# Rethinking AI in Medical Imaging: Assumptions, Reality, and Reframing; Towards Physician-Centered Agentic Reasoning

Arman Rahmim*[1,2,3], Nourhan Bayasi[1,2], Xiaoxiao Li[4,5], Babak Saboury[1,6], Fereshteh Yousefirizi[1,2]

[1] Department of Basic and Translational Research, BC Cancer Research Institute, Vancouver, BC, Canada
[2] Department of Radiology, University of British Columbia, Vancouver, BC, Canada
[3] Departments of Physics and Biomedical Engineering, University of British Columbia, Vancouver, BC, Canada
[4] Department of Electrical and Computer Engineering, University of British Columbia, Vancouver, BC, Canada
[5] Vector Institute, Canada
[6] Institute of Nuclear Medicine, Bethesda, MD, USA

## Abstract

Medical imaging has served as primary proving ground for clinical AI, yet a decade of intense research has not translated into proportionate bedside impact. We argue that this gap is not primarily a product of insufficient algorithmic performance, inadequate regulation, or limited explainability. Rather, it reflects a structural misalignment, between how AI systems are designed and evaluated, and how clinical decisions are made. This work identifies 8 interconnected dimensions of this misalignment: 1) the dominance of pixel-only models in a multimodal clinical world; 2) the erosion of physician trust through opaque and inflexible systems; 3) the unfulfilled promise of foundation models in data-sparse medical domains; 4) the persistent bottleneck of non-shareable under-curated datasets; 5) the gap between validated algorithms and deployable clinical platforms; 6) the slow, unmonitored translation of evidence into routine practice; 7) the risk of AI displacing rather than extending physician judgment; and 8) the failure of prediction-centric AI to generate actionable, agentic clinical guidance. For each dimension, we reframe the problem and propose a path forward, culminating in a vision of physician-centered AI solutions with agentic reasoning that extend, rather than replace, clinical judgment.

*Corresponding author: arman.rahmim@ubc.ca

## Introduction

Artificial intelligence (AI) in medical imaging has undergone rapid expansion, with proposed solutions proliferating across nearly every domain of clinical practice. The field, however, is replete with exaggerated claims (whether of the present or of the future) [1]. Less than a decade ago, convolutional neural networks were confidently predicted to replace radiologists, a prediction that, in hindsight, failed due to misunderstanding what radiologists do and underestimating technical constraints; and today, foundation models are promoted as universal engines of medical reasoning [2]. Clinical reality, however, remains sobering: adoption is patchy, impact modest, and trust fragile [3]. The widening gap between technical breakthroughs and bedside benefit is not due to regulatory inertia or clinician caution alone; it reflects deeper misalignments between how AI is developed and how medicine is practiced. Medical imaging exposes these tensions most clearly. Here, clinicians do not reason from pixels alone. They integrate laboratory trends, comorbidities, therapies received, prior imaging, and evolving disease trajectories, particularly in complex or ambiguous cases. AI algorithms designed around images alone, however elegant, may fall short of clinical relevance in complex, context-dependent scenarios. Broader challenges persist in models that collapse patients into voxel grids, tools that fail to earn physician trust, foundation models that scale without context, and a persistent scarcity of curated multimodal datasets. This paper does not claim to introduce an entirely new problem; rather, it offers an integrated perspective across these interconnected dimensions, oriented toward embedding AI as an extension of the physician and grounded in the realities of clinical practice. We argue that these are not independent technical shortcomings but expressions of a single structural problem: medical imaging AI has been optimized for benchmark performance rather than clinical utility. Solving this requires not incremental algorithmic refinements, but fundamental reorientation, from AI as a product to AI as a physician-aligned capability embedded in the realities of clinical practice.

These tensions between technological assumptions and clinical reality occur across nearly every dimension of medical imaging AI [4]. They are not isolated problems but manifestations of a deeper structural mismatch in how AI is conceived, evaluated, and deployed. This broader trajectory is captured in Fig. 1, which maps 8 major themes in the field, spanning clinical context, trust, foundation models, data, deployment platforms, implementation science, and agentic AI, across three stages: early optimism, practical constraints, and clinically grounded reframing; illustrating the shift from model-centric innovation toward clinically integrated, context-aware, physician-aligned AI. A growing body of literature has examined aspects of this misalignment, including issues of workflow integration, model generalizability, and clinical trust. The present work seeks to contribute a synthesis across these interdependent aspects, spanning clinical trust, multimodal data integration, implementation science, and agentic AI, brought together within a single clinical alignment framework. We acknowledge that several themes discussed here, particularly data scarcity, harmonized sharing, and governance, are themselves technical and policy challenges; the intent is not to avoid those domains but to situate them within a broader clinical alignment framework.

***1) Beyond pixels.*** Much of imaging AI has inherited a paradigm from natural image analysis, reducing patients to image voxel arrays and diagnostic problems to classification tasks. Well-validated image-only AI systems, such as large vessel occlusion triage tools and breast cancer screening algorithms, already demonstrate meaningful clinical value. Radiologists often function effectively even with limited clinical history, and in many settings, this remains the norm. Nevertheless, for complex or ambiguous cases, the integration of laboratory values, comorbidities, prior therapies, and treatment histories has potential to substantially enhance diagnostic accuracy and clinical utility. Models designed with no ability to access or incorporate such context may be technically sophisticated yet constrained in their applicability to nuanced clinical scenarios. As an example, variations in patient hemoglobin levels during prostate cancer therapies are significant considerations for subsequent therapy plans [5] , and an AI-based clinical decision support (CDS) tool that only relies on radiological images will be extremely limited in such contexts. As such, AI models that assess tumor response without access to concurrent laboratory trends or systemic therapy records may mischaracterize disease trajectory in ways that image data alone cannot resolve [6].

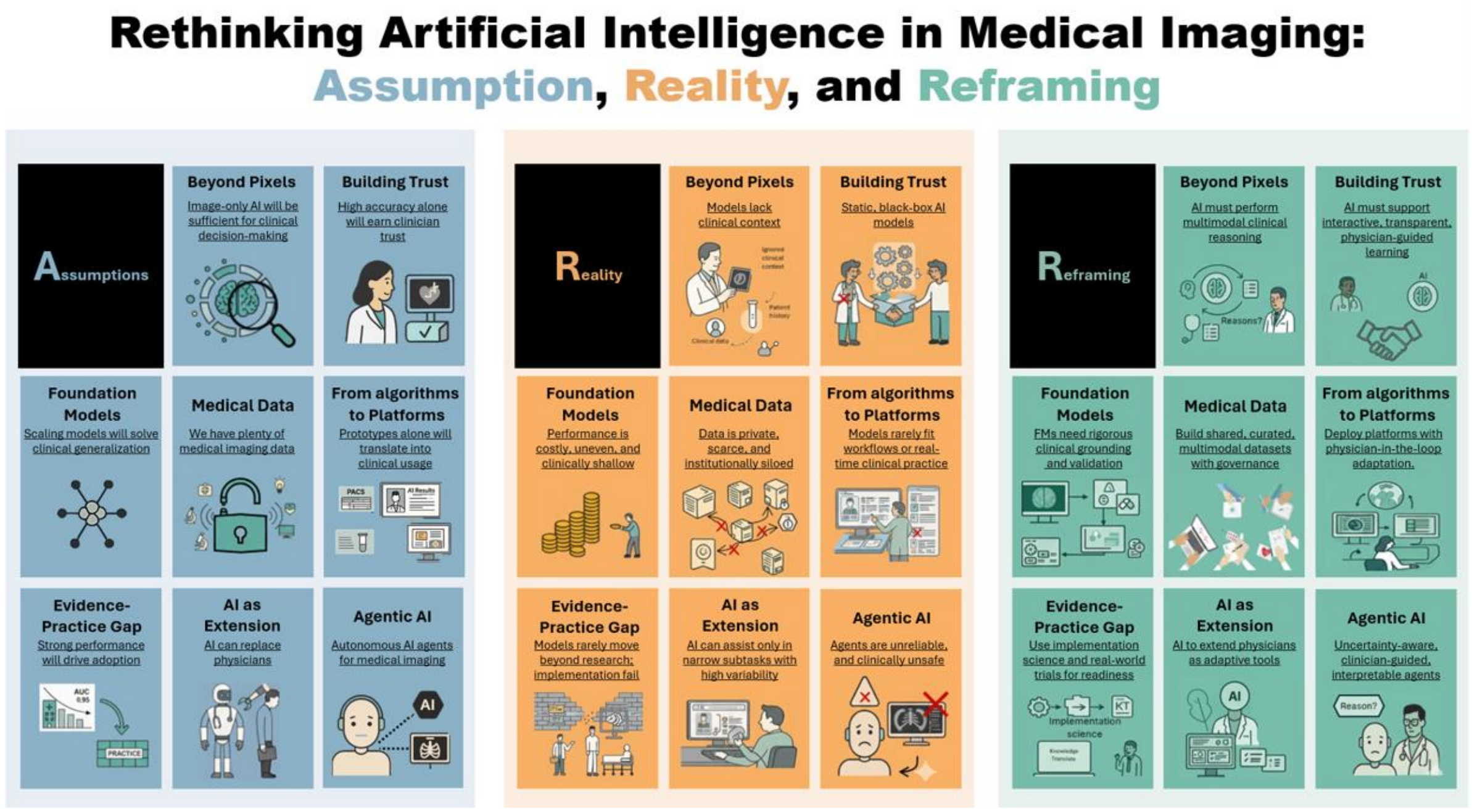

Figure 1: Overview of the evolving landscape of AI in medical imaging, spanning 8 interconnected dimensions: models that go beyond pixels, physician trust, foundation models, medical data, the path from algorithms to platforms, the evidence–practice gap, AI as an extension of the physician, and agentic AI. Each dimension is shown across three stages: Assumptions (what the field has assumed or believed), Reality (what is actually observed or what has failed), and Reframing (what is needed for meaningful clinical impact). Together, these trajectories illustrate the broader shift from model-centric innovation to clinically integrated, context-aware, physician-aligned AI.

***2) Building trust.*** Trust cannot be legislated into existence; it emerges from a more intricate interplay of utility, reliability, transparency, alignment of values, and professional agency [7]. Current AI systems too often present themselves as static "black box" solutions with middling accuracy, leaving clinicians frustrated [8]. In the physician-in-the-loop AI model, the clinician acts as an active participant who labels data, corrects outputs, and steers model learning, with the primary goal of improving AI performance over time. In the AI-in-the-loop physician model, by contrast, AI serves as a continuous decision-support layer embedded in the clinician's workflow: surfacing relevant information, flagging uncertainty, and offering evidence-grounded recommendations, while the physician retains full authority over the final decision. The first model centers on improving the AI iteratively through human correction; the second centers on augmenting human judgment in real time without requiring the physician to act as a data labeler.

Importantly, neither model can function without addressing the realities of clinical data. In medical imaging, we rarely have the scale, heterogeneity, and longitudinal diversity needed to train models that can internalize the full breadth of clinical variability. By contrast, physicians accumulate thousands of hours of exposure to rare presentations, edge cases, artifacts, and multimodal clinical context, as forms of experiential knowledge that current models do not replicate and that AI systems must increasingly seek to incorporate through active and continual learning frameworks [9–12]. Such frameworks allow AI tools to evolve while avoiding catastrophic forgetting.

Trust also depends on how AI systems present themselves to individual clinicians. Clinicians differ in their tolerance for errors, decision boundaries, and workflow priorities [13,14]. Rather than mirroring idiosyncratic interpretive tendencies, which would encode inter-reader variability as a feature rather than addressing it as a known diagnostic quality challenge, AI systems should aim to standardize evidence-based pathways while remaining adaptable to different clinical workflow contexts and communication preferences. Interfaces must empower clinicians through intuitive interactions, such as detection modification and segmentation refinement. Only when AI systems behave as adaptable, transparent, and responsive partners, not distant authorities, can they earn a place as trusted colleagues in clinical practice.

Three concrete gaps limit how current physician-in-the-loop AI interacts with and incorporates physician input. First, repetitive annotation burden: physicians routinely re-draw contours or re-click boxes on every case, even every lesion, adding hands-on time to an already time-pressured read. Second, error repetition: a standard model corrects an error once and forgets it, repeating the same mistake case after case because it lacks any mechanism to carry correction patterns forward. Third, poor uncertainty communication: physicians are rarely told, case by case, how reliable the AI's output is, or when their scrutiny is most likely to add value. Agentic reasoning addresses all three: by carrying correction patterns forward, calibrating uncertainty case by case, and reducing annotation burden through learned physician-specific thresholds. That these gaps matter clinically is underscored by recent evidence: a 2024 meta-analysis of 106 studies found that human–AI pairs perform significantly worse, on average, than the better of the human or AI alone, with the sharpest losses in decision-making tasks, and a randomized trial showed no diagnostic improvement when physicians were supported by a query-only LLM even though the LLM alone outperformed them. Pairing can outperform either party when physicians have something to actively interrogate rather than a pre-computed output to accept or override out of habit, which demands appropriate reliance: the ability to discern when the AI is likely correct and to act on that judgment rather than defer uniformly.

***3) Promise and limits of foundation models.*** Foundation models have been celebrated for their ability to learn general-purpose representations from large datasets. In practice, however, gains in medical imaging have been modest, task-dependent and inconsistent across institutions [15,16]. Their energy demands are substantial, raising questions about sustainability of scaling as a primary strategy [17,18]. The argument here is not against scale per se, but against scale as a substitute for clinical grounding: without corresponding improvements in data quality, physician-driven validation, and evaluation rigor, larger models are unlikely to close the gap between research performance and real-world impact. Notably, the field has increasingly embraced self-supervised and weakly supervised learning approaches that reduce dependence on curated labels; a shift that partially addresses the labeling bottleneck, though it does not eliminate the need for careful validation and clinical alignment.

In data-sparse domains like medical imaging (and specific modalities within it), genuine progress will depend on carefully curated multimodal datasets paired with rigorous evaluations of clinical benefit [19]. Curation, however, is not synonymous with sanitization: models trained only on clean, carefully selected data may fail to generalize the noise and heterogeneity of routine clinical practice. Curation efforts should therefore preserve representative diversity rather than eliminate it. The goal is not simply more data, but better-structured data aligned with the decisions clinicians face. Without these, foundation models may shine in theory yet deliver only modest returns where it matters most: the bedside.

***4) A central bottleneck in data.*** Medical imaging continues to face profound data scarcity, not only a lack of large, labeled cohorts but, more importantly, a lack of shareable datasets. A 2022 meta-research study found that fewer than 2% (4 of 218) of radiology AI articles published from 2017–2021 shared both code and complete experimental data, in stark contrast to the 40–50% code-sharing rates seen in general computer vision and NLP research [20]. Without broad collaboration, harmonized data sharing, and structured generation of multimodal imaging resources, AI advances in medicine will remain incremental.

A practical way to mitigate limited shareable cohorts is to expand data diversity via clinically grounded augmentation and constrained synthetic generation. Basic perturbations improve robustness but do not create the rare anatomies, edge-case disease patterns, and institution-specific acquisition signatures that drive real-world failure. Generative models can target these missing modes, but only if synthesis is conditioned on clinically meaningful factors and screened for anatomical plausibility and label consistency; otherwise, augmentation quietly injects hallucinated pathology and unrealistic features. A reliable strategy is controlled generation plus automated filtering and clinician spot-audits, prioritizing under-represented subgroups and hard cases. Another often-overlooked path is revisiting the ethics and governance of medical data sharing. It has been argued [21,22] that patients do not "own" their medical data; instead, institutions should act as custodians responsible for stewardship and socially beneficial use. This perspective supports the creation of more permissive, standardized, and well-governed data-access frameworks, essential prerequisites for reproducible, multicenter AI research and for breaking the data bottleneck limiting clinical translation.

***5) From algorithms to platforms.*** Algorithms, no matter how advanced, cannot transform practice into isolation. Translation requires platforms: integrated systems that plug into daily workflows, connect with hospital PACS, and evolve iteratively with clinician input. Such platforms must address the full integration challenge: bridging AI models with electronic health records, handling diverse imaging protocols across scanners and institutions, supporting human-in-the-loop feedback loops, and enabling deployment at scale without sacrificing safety or interpretability. Emerging open frameworks illustrate what clinically integrated platforms can look like in practice. MONAI Label, for instance, enables active learning workflows by connecting AI-assisted annotation tools directly with clinical annotation pipelines, reducing labeling burden while continuously refining model performance. Kaapana provides an open-source, container-based platform for deploying AI pipelines in federated, privacy-preserving clinical environments, enabling multi-institutional collaboration without centralizing sensitive data. These frameworks represent a meaningful step toward making AI deployment a practical, maintainable endeavor rather than a one-time research artifact. Realizing this philosophy at scale will require not only technical infrastructure but governance frameworks and reimbursement models that support sustainable clinical AI deployment as an ongoing institutional commitment.

***6) Closing the evidence–practice gap.*** Even when strong evidence exists for the value of a particular solution, implementation is not guaranteed. Historically, effective interventions have taken 15–17 years to reach routine practice [23,24], and there is little reason to assume AI will be exempt from this inertia without deliberate strategies to accelerate translation. Implementation science, which is a systematic study of methods to promote the uptake of evidence-based interventions into routine practice, offers a structured path forward [25,26]. Its tools include identifying context-specific barriers to adoption, engaging stakeholders early, and designing interventions that fit existing workflows rather than demanding that workflows adapt to them.

For medical imaging AI specifically, this means deploying integrated knowledge translation (iKT) approaches: co-creating AI solutions with physicians from the initial definition of clinical problems through to the design and execution of evaluation studies. Clinicians who participate in defining the problem are far more likely to recognize the solution as meaningful, to flag implementation barriers early, and to advocate for adoption within their institutions. Without this co-design process, even technically validated models' risk being solutions to problems clinicians did not ask for, evaluated on metrics clinicians do not use, deployed in workflows clinicians were not consulted about.

The evidence–practice gap in medical imaging AI is also widened by a monitoring problem that implementation science has begun to address in other domains but that remains largely unsolved here: most health systems currently lack the infrastructure to detect when deployed AI tools begin to degrade, whether due to shifts in patient populations, changes in imaging protocols, or scanner upgrades. Closing this gap requires not only successful initial deployment but sustained institutional commitment to post-deployment surveillance, re-validation, and iterative improvement.

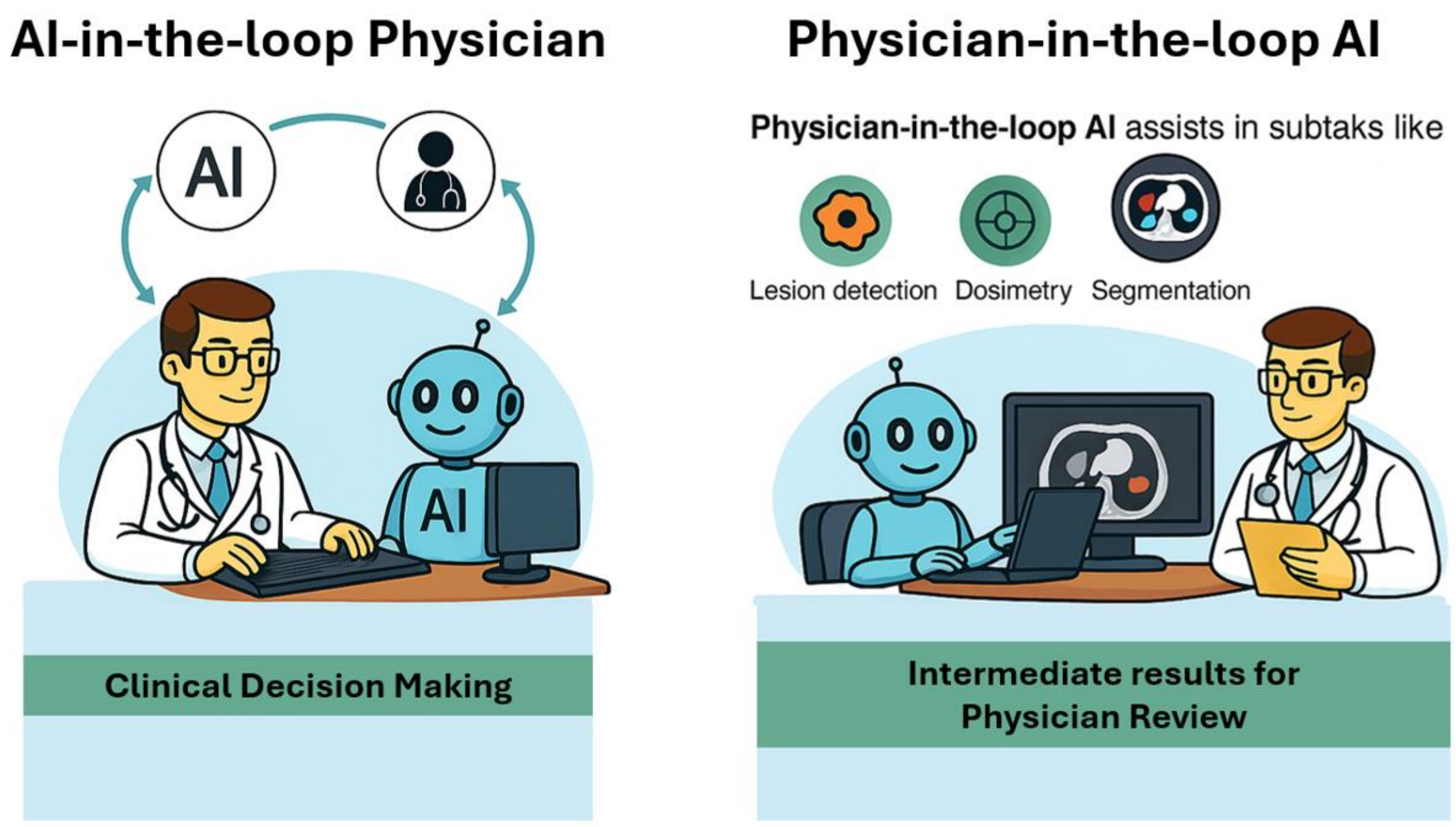

Figure 2: Empowering physicians to make more autonomous, informed and effective decisions; inter-linking of physician-in-the-loop AI and AI-in-the-loop physician.

***7) AI as extension of the physician.*** The goal is not to replace physicians but to extend their capabilities; e.g., augmenting what they can perceive, synthesize, and act upon, rather than substituting for the judgment that remains irreducibly theirs. For subtasks such as segmentation, dosimetry and lesion detection, physician-in-the-loop AI can improve reproducibility and efficiency. For clinical decision-making, physicians remain central. The true vision is AI-in-the-loop physicians: clinicians empowered by AI to make more informed, autonomous and effective decisions, a physician equipped with AI-enabled armor rather than replaced by it (Fig. 2). The distinction between these two modes, elaborated in the Building Trust section above, is operationally meaningful: the former improves the model iteratively via human correction; the latter improves the clinician's decision-making in real time via AI-generated insights, without requiring the physician to act as a data labeler.

Realizing this vision requires AI that adapts to how the individual physician works. Physician behavior is measurable, through cursor movements, dwell times, report edits, confidence ratings, and gaze patterns, and these signals can be used to build physician-specific models of reasoning. Preliminary work has demonstrated that an agent trained on a physician's history of accepted, revised, and overridden segmentation suggestions can anticipate the type and location of the next correction, significantly cutting interactions while improving accuracy (see last section). This personalization must be robust: when insufficient data exists for a given physician on a given task, the system should default to a shared non-personalized model, and the balance between the two should shift transparently as data accumulates. The goal is not a single fixed interaction style imposed on all clinicians, but an AI collaborator that becomes recognizably adapted to each physician, knowing when to propose, when to hold back, and how much explanation to provide.

***8) Actionable and Agentic AI: aligning imaging predictions with clinical decisions.*** This dimension concerns the failure of prediction-centric AI to generate actionable clinical guidance, and the agentic architectures needed to close this gap. A recurring limitation of medical imaging AI is not its ability to predict, but its failure to translate those predictions into concrete clinical action [27]. Clinicians do not primarily ask whether an outcome can be predicted; they inquire as to what should change in patient management, when a change should occur, and perhaps how this prediction happened (explainability). Risk estimates that are not clearly linked to a specific clinical choice, such as adjusting follow-up strategies, modifying treatment assessments, or altering the timing of intervention, remain abstract and difficult to use in practice.

Actionable AI must therefore be built around decisions rather than predictions. Models should be designed to answer clinically meaningful questions, aligned with relevant time horizons, acceptable levels of uncertainty, and the consequences of acting versus not acting [28]. This shift requires moving beyond generic performance metrics toward evaluation frameworks that reflect how model outputs influence different clinical strategies (task-based evaluation) [8,29,30]. Without this alignment, even well-validated models may produce results that are technically correct yet offer little practical value at the bedside. Embedding actionability from the outset requires early co-definition of use cases with clinicians, explicit consideration of decision trade-offs, and interfaces that allow clinicians to interpret, question, and adapt outputs within the context of individualized patient care.

## Towards physician-centered AI solutions with Agentic Reasoning

To bridge the gap between pixel-level accuracy and patient-level care, the field must explore agentic AI solutions. Traditional deep learning treats diagnosis as a static, one-shot classification task, brittle when faced with ambiguity and blind to clinical history. Agentic systems, by contrast, solve the missing context problem by actively orchestrating the diagnostic pipeline. Instead of passively accepting an input image, an agent can reason dynamically: querying the Electronic Health Record (EHR) for lab trends to resolve a radiologic uncertainty, requesting alternative views, or cross-referencing prior reports. Community efforts such as MedAgentBench [31], illustrate a shift toward self-improving, human-in-the-loop workflows that go beyond passive prediction to enable context-aware planning, iterative reflection, and adaptive action.

We propose that agentic architectures in medical imaging should be conceived around three functional capacities that traditional models lack. First, context orchestration: the ability to autonomously retrieve and synthesize relevant clinical information from EHRs, prior imaging, laboratory systems, and genomic databases before generating a recommendation. Second, uncertainty-aware reasoning: the capacity to recognize the limits of its own competence, flag cases that fall outside its reliable operating range, and escalate appropriately to human review rather than producing confident but unreliable outputs. Third, action-oriented output: producing not just predictions or probabilities, but structured recommendations linked to specific decision points, such as "recommend biopsy given PSA trajectory and lesion characteristics" rather than "probability of malignancy: 74%". This shift moves AI from "System 1" pattern matching to "System 2" deliberative reasoning. Yet significant challenges remain. Current systems lack robust mechanisms for recognizing the boundaries of their competence, and autonomous actions introduce new accountability concerns that existing regulatory frameworks are not yet equipped to address. Regulatory sandboxes and staged deployment models, analogous to Phase II/III trial logic but adapted for adaptive AI systems, will be needed to generate the post-market evidence required for safe scaling. Agentic AI should be understood not as autonomous replacement of clinical reasoning, but as a structured capability augmentation: a system that meets the physician inside their workflow, brings the relevant information to bear, and flags what it does not know, leaving final judgment unambiguously with the clinician. This is the architecture of AI that medicine needs.

These capacities are not merely theoretical. Recent agentic segmentation systems illustrate how each can be realized concretely in an imaging pipeline, addressing 3 significant existing gaps, as also depicted in Fig. 3:

(i) *Repetitive annotation burden:* A physician who must repeatedly re-annotate the same case until an output is acceptable faces a real time and effort burden; Clinical-SAM [32] instead trains a frozen segmentation agent to learn an individual physician's acceptance standard from nothing more than binary accept/reject feedback, closing this loop without additional annotation [33].

(ii) *Repeated AI mistakes:* Without carried-forward context, the same errors can recur across cases; LORE [34] maintains a memory bank of prior physician–image interactions so that a correction made once is applied automatically on subsequent cases, rather than discarded [35].

(iii) *Uncalibrated uncertainty:* Where conventional outputs give no indication of their own reliability, PILOT [36] reports calibrated confidence alongside each proposed output and suggests the appropriate next action, operationalizing uncertainty-aware reasoning directly at the point of decision.

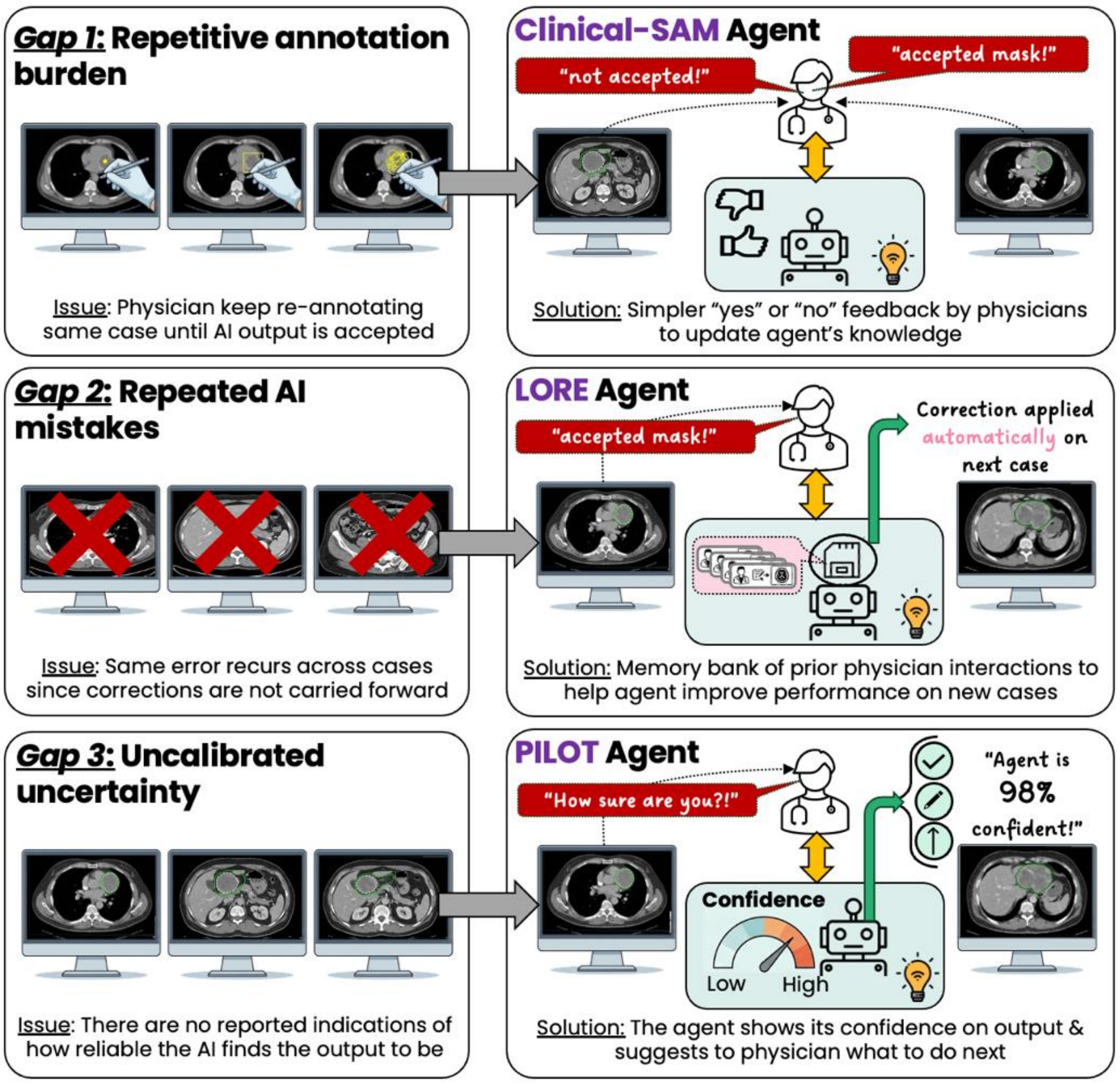

Figure 3: Three practical gaps limiting physician-facing agentic AI, and how minimal-feedback agentic systems close them. *(i) Repetitive annotation burden*: physicians re-annotate the same case until an AI output is accepted; Clinical-SAM instead learns a physician's individual acceptance standard from simple accept/reject feedback. *(ii) Repeated AI mistakes*: without memory, the same error recurs across cases; LORE maintains a memory bank of prior physician–image interactions so corrections carry forward automatically. *(iii) Uncalibrated uncertainty:* conventional outputs give no indication of their own reliability; PILOT reports calibrated confidence alongside each output and recommends the appropriate next step.

## Conclusion

The eight reframings advanced in this work converge on a single design imperative: medical imaging AI must be built for the physician, not merely handed to the physician. This means co-designing systems with clinicians from problem definition to deployment; measuring success by decision quality and patient outcomes, not benchmark accuracy; building platforms that integrate into workflows rather than disrupting them; and designing agentic systems that bring context, reasoning, and appropriate humility to every clinical encounter. None of this is technically impossible, frameworks like MONAI, Kaapana, and emerging agentic benchmarks provide concrete starting points. What is missing is not capability but orientation. The field has spent a decade asking, "can AI match physician performance on task X?" The more important question is "does AI make physicians better at caring for patients?" Answering this question requires sustained collaboration across machine learning, clinical medicine, implementation science, and health policy, and a shared commitment to measuring progress not by papers published but by patients helped.

### Conflicts of Interest

Dr. Arman Rahmim is a co-founder of Ascinta Technologies. Dr. Babak Saboury is co-founder of United Theranostics. All authors declare no relevant conflicts of interest.

### Acknowledgments

We acknowledge support from the Natural Sciences and Engineering Research Council of Canada (NSERC) Discovery Horizons Grant DH-2025-00119, and the BC Cancer Foundation.